\documentclass[12pt,a4paper]{article}
\pdfoutput=1
\usepackage{amsmath,amsfonts,amssymb}
\usepackage{graphicx}
\usepackage{bbm}
\usepackage{xcolor}
\usepackage{cite}
\usepackage{url}
\usepackage{hyperref}

\newcommand{\be}{\begin{equation}}
\newcommand{\ee}{\end{equation}}
\newcommand{\bea}{\begin{eqnarray}}
\newcommand{\eea}{\end{eqnarray}}
\newcommand{\besub}{\begin{subequations}}
\newcommand{\eesub}{\end{subequations}}
\newcommand{\ba}{\begin{array}}
\newcommand{\ea}{\end{array}}
\newcommand{\bi}{\begin{itemize}}
\newcommand{\ei}{\end{itemize}}
\newcommand{\nn}{\nonumber}

\newcommand{\GeV}{{\rm GeV}}
\newcommand{\TeV}{{\rm TeV}}

\newcommand{\sw}{{s_W^2 }}

\newcommand{\duLR}{{\ensuremath{(\delta^u_{23})_{LR}}}}
\newcommand{\duLL}{{\ensuremath{(\delta^u_{23})_{LL}}}}
\newcommand{\ddLR}{{\ensuremath{(\delta^d_{23})_{LR}}}}
\newcommand{\ddLL}{{\ensuremath{(\delta^d_{23})_{LL}}}}

\newcommand \vect[1]{\left(\begin{matrix} #1 \end{matrix}\right)}

\begin{document}

\setlength\arraycolsep{2pt}

\setcounter{page}{1}

\setlength\arraycolsep{2pt}

\begin{titlepage}  

\rightline{\footnotesize{DO-TH 12/01}} \vspace{-0.2cm}

\begin{center}

\vskip 0.4 cm

{\Large  \bf Squark Flavor Implications from $\bar B \to \bar K^{(*)} l^+ l^-$}

\vskip 0.7cm

{\large
Arnd Behring$^{a}$, Christian Gross$^{a,b}$, Gudrun Hiller$^{a}$ and Stefan Schacht$^{a}$
}

\vskip 0.5cm

{\it
$^{a}$Institut f\"ur Physik, Technische Universit\"at Dortmund, D-44221 Dortmund, Germany\\
$^{b}$Departement Physik, Universit\"at Basel, CH-4056 Basel, Switzerland
}

\vskip 1cm

\end{center}

\begin{abstract}
Recent experimental and theoretical progress regarding $\bar B \to \bar K^{(*)} l^+ l^-$ decays led to improved bounds on the Wilson coefficients  $C_9$ and $C_{10}$ of four-fermion operators
of the $|\Delta B|=|\Delta S|=1$ effective  Hamiltonian. 
We analyze the resulting implications on squark flavor violation in the MSSM and
obtain new constraints on 
flavor-changing left-right mixing in the up-squark-sector. 
We find  the dimensionless flavor mixing parameter $(\delta^u_{23})_{LR}$, depending on the flavor-diagonal MSSM masses and couplings,  to be as low as  $\lesssim 0.1$.
This has implications for models based on radiative flavor violation
and  leads to ${\cal{B}}(\bar B_s \to \mu^+ \mu^-) \gtrsim 1 \times 10^{-9}$.
Rare top decays  $t \to c \gamma, t \to c g, t \to c Z$ have branching ratios predicted to be below $\lesssim  \mbox{few} \times 10^{-8},  10^{-6}$
and $10^{-7}$, respectively.
\end{abstract}

\end{titlepage}

\newpage

%%%%%%%%%%%%%%%%%%%%%%%%%%%%%%%%%%%%%%%%%%%%
\section{Introduction} 

Heavy flavor physics is rapidly advancing with the successful start of the Large Hadron Collider's (LHC)
$b$-physics program and the final analyses from the Tevatron as well as the
$B$-factory experiments Belle and BaBar.
Most notably the LHCb collaboration is currently making a clean sweep in model space around the
Standard Model (SM): The current upper limit
on the $\bar B_s \to \mu^+ \mu^-$ branching ratio at 95\%  (90\%) C.L.  \cite{Aaij:2012ac}
\begin{align} \label{eq:bsmumulhcb}
{\cal{B}}(\bar B_s \to \mu^+ \mu^-) < 4.5 \, (3.8) \times 10^{-9}
\end{align}
is down to  the level of the SM, 
$ {\cal{B}}(\bar{B}_s \to \mu^+\mu^-)_{\rm SM} \sim (2.5-4) \times 10^{-9}$,  
e.g., \cite{Buchalla:2008jp,Hiller:2011sg}. The SM prediction is much more precise
assuming the measured $B_s -\bar B_s$ mass difference to be SM-like,
$ {\cal{B}}(\bar{B}_s \to \mu^+\mu^-)_{\rm SM} = (3.2 \pm 0.2) \times 10^{-9}$  \cite{Buras:2011fz}.
Another milestone constitutes the preliminary measurement  of 
 the position of the zero of the forward-backward asymmetry
in $\bar B \to \bar K^{0*} \mu^+ \mu^-$ decays \cite{LHCbMoriond2012}
\begin{align} \label{eq:zeroLHCB}
q_0^2 =4.9 ^{+1.1}_{-1.3} \, \mbox{GeV}^2,
\end{align}
consistent with the SM prediction $q_0^2|_{\rm SM}= 4.0 \pm 0.3 \, \mbox{GeV}^2$ \cite{Bobeth:2011nj}, also\cite{Beneke:2004dp,Ali:2006ew}.

In this work we aim at investigating the  space for supersymmetric  flavor physics in the light of the recent and new data from direct collider searches and on  flavor-changing rare processes 
$b \to s \, l^+ l^-$. Especially, the study of the exclusive modes
$\bar B \to \bar K^{(*)} \mu^+ \mu^-$, which are accessible to hadron colliders, has progressed
significantly over the last year(s) both experimentally and theoretically. The latter is due to the exploitation
of the region where the invariant mass of the dilepton is large, of the order of the $b$-quark mass
~\cite{Bobeth:2010wg}, by making use of the heavy quark effective theory framework of Ref.~\cite{Grinstein:2004vb}.
Using the recent data from CDF \cite{Aaltonen:2011ja} and LHCb~\cite{LHCbdata}, 
improved model-independent constraints on the Wilson coefficients $C_{9,10}$ of the semileptonic four-fermion operators $O_9$ and $O_{10}$ have been obtained~\cite{Altmannshofer:2011gn,Bobeth:2011nj}.

Here, we study the implications of these $C_{9,10}$-constraints on squark flavor violation in the minimal supersymmetric SM (MSSM).  Flavor violation in
supersymmetric (SUSY) models originates from generational mixings in the sfermion mass matrices.  One-loop SUSY effects including their phenomenology in semileptonic $B$-decays are known for a while~\cite{Cho:1996we,Hewett:1996ct,Lunghi:1999uk,Ali:2002jg}.
We work out  the implications of the improved data on  semileptonic decays,
taking into account the recent direct search limits on SUSY particles.

The plan of this paper is as follows:
In Section~\ref{sec:squark_flavor} we explain, after introducing the $|\Delta B|=1$ effective theory  framework, why only chirality-flipping flavor mixing between the second and third generation in the up-sector, parametrized as $\duLR$, has a significant sensitivity to the 
$C_{9,10}$-constraints.
We then describe the method of SUSY parameter scanning and discuss the constraints on 
$\duLR$.
Results  are given in Section \ref{sec:predictions}, including predictions for $b$-physics,
implications for the models with radiative flavor violation (RFV)  recently discussed in Refs.~\cite{Crivellin:2008mq, Crivellin:2011sj}, and  rare top decays.
 We summarize in Section~\ref{sec:conclusions}. Details on the SUSY loop contributions to $b \to s l^+l^-$ processes including corrections to the literature are given in the Appendix.

%%%%%%%%%%%%%%%%%%%%%%%%%%%%%%%%%%%%%%%%%%%%
\section{Constraining squark flavor mixing} 
\label{sec:squark_flavor}

In Section \ref{sec:Heff}  the effective theory for $|\Delta B|=1$ transitions is described.
In Section  \ref{sec:sflavorinraredecays} we discuss which flavor parameters in supersymmetric models can be effectively probed in semileptonic decays. The constraints are worked out in Section \ref{sec:numerical}, and discussed in Section \ref{sec:bounds}.

\subsection{The \texorpdfstring{$|\Delta B|=1$}{|Delta B| = 1} effective theory framework \label{sec:Heff}}

In order to describe rare decays of $B$-mesons, we employ the effective Hamiltonian
\begin{align} \label{eq:Heff}
\mathcal{H}_{\rm eff} &= -\frac{4 G_F}{\sqrt{2}} V_{tb} V_{ts}^* \sum_{i} C_i(\mu) O_i(\mu) + \textrm{h.c.} \,,
\end{align}
where the $C_i$ are Wilson coefficients and the $O_i$  higher-dimensional $\vert\Delta B\vert=\vert\Delta S\vert=1$ operators, and
$V$ denotes the Cabibbo-Kobayashi-Maskawa (CKM) matrix, $G_F$  the Fermi constant and $\mu$  the factorization scale.
The New Physics (NP)  contribution to the current-current ($O_{1,2}$), QCD penguin ($O_{3-6}$) and chromomagnetic dipole ($O_{8}$) 
operators can be neglected here (cf., e.g.,~\cite{Cho:1996we}). 
The most relevant operators are the electromagnetic dipole operator $O_{7}$ and the semileptonic  
four-fermion operators $O_{9,10}$, written as
\begin{align}
O_7 &= \frac{e}{16\pi^2} m_b \left(\bar{s}_{L} \sigma_{\mu\nu}  b_{R} \right) F^{\mu\nu} \, ,& 
 \nn \\
O_9 &= \frac{e^2}{16\pi^2} \left(\bar{s}_{L} \gamma_\mu  b_{L}\right) \left(\bar{l} \gamma^\mu l\right) 
\, ,&
O_{10} &= \frac{e^2}{16\pi^2} \left(\bar{s}_{L} \gamma_\mu  b_{L} \right) \left(\bar{l} \gamma^\mu \gamma_5 l \right).  \label{eq:ops}
\end{align}
The $\overline{\mathrm{MS}}$ mass of the $b$-quark is denoted by $m_b$, and we neglect
the mass of the strange quark.

In general models for NP, the effective Hamiltonian contains additional operators with flipped chirality.
In the SM and in minimally flavor-violating (MFV) SUSY models with CKM-induced flavor violation, these are suppressed by $m_s/m_b$ compared to the corresponding unflipped operators given in Eq.~(\ref{eq:ops}). The non-MFV chargino contributions to $b \to s l^+ l^-$ processes share this feature of suppressed chirality-flipped contributions.
As discussed below, we will  be mainly interested in chargino loop-induced contributions, so it is justified to neglect the chirality-flipped operators and  consider left-handed currents only. 
Furthermore, 
we do not take into account the effects of the scalar- 
and pseudoscalar semileptonic operators  (and their chirality-flipped counterparts). This  is a good approximation for not too large $\tan\beta$~\cite{Carena:2000uj}, to which we restrict our analysis.
The exclusion of a sizable enhancement in the $\bar B_s \to \mu^+ \mu^-$ branching ratio,
see Eq.~(\ref{eq:bsmumulhcb}), supports this further.

The NP contribution, which in our case is the MSSM contribution, can be split into  one
stemming from the diagonal elements of the squark mass matrices, labeled 'diag', and the 
remainder induced by 
the corresponding flavor off-diagonal entries as
\begin{align}
C_i^{\mathrm{NP}} &= C_i^{\textrm{diag}}+C_i^{\textrm{MI}} \,,
\end{align}
where $C_i^{\mathrm{NP}}$ denotes the NP contribution to the $C_i$, i.e., $C_i = C_i^{\mathrm{SM}} + C_i^{\mathrm{NP}} $.  We use the terminology `mass insertion' (MI)  since we will constrain the commonly used MI parameters, i.e., the off-diagonal elements of the squark mass term divided by an average squark mass squared. 
Note, however, that we do not rely on the MI approximation ~\cite{Hall:1985dx,Gabbiani:1988rb} in the numerical analysis, but instead use the exact formulae.

We employ the two-loop matching conditions of Ref.~\cite{Bobeth:1999mk} for the SM contribution and the one-loop results of~\cite{Cho:1996we,Lunghi:1999uk,Hewett:1996ct} for the MSSM contribution. The Wilson coefficients at the relevant scale for $b$-decays, $\mu_b={\cal{O}}(m_b)$, are obtained from the ones at the matching scale $\mu_0$ of ${\cal{O}}(100 \, \mbox{GeV})$ by solving the RG equations. We perform this by extending and using
the flavor tool \texttt{EOS}~\cite{eos}.
The values of the SM parameters used are given in Table \ref{tab:SM}.
The SM values of the most important Wilson coefficients are given as (at $\mu_b=4.2$ GeV)
\begin{align} \label{eq:SMvalues}
C_7^{\rm SM} (\mu_b)   &= -0.33, &
C_9^{\rm SM}(\mu_b)    &=  4.27, &
C_{10}^{\rm SM}(\mu_b) &= -4.15\, .
\end{align}
\begin{table}
\begin{center}
 \begin{tabular}{|l|ll|} \hline
$m_t^{\mathrm{pole}}$		& $173.3 \, \mathrm{GeV}$ &  \cite{:1900yx}\\  
$m_b(m_b)$	& $4.19   \, \mathrm{GeV}$ &  \cite{Nakamura:2010zzi} \\
$m_W$					& $80.399 \, \mathrm{GeV}$ &  \cite{Nakamura:2010zzi} \\
$m_Z$					& $91.1876  \, \mathrm{GeV}$ &  \cite{Nakamura:2010zzi} \\
$\alpha_s(m_Z)$			& $0.1184$ & \cite{Nakamura:2010zzi}  \\
$s_W^2$		& $0.23116$  &  \cite{Nakamura:2010zzi} \\
\hline
\end{tabular}
\caption{The values of the Standard Model parameters used in this work. \label{tab:SM}}
\end{center}
\end{table}

\subsection{SUSY flavor contributions to \texorpdfstring{$b \to s l^+ l^-$}{b -> sl+l-} \label{sec:sflavorinraredecays}}

We begin  by discussing which MI parameters have the best potential of receiving significantly improved constraints from the $C_{9,10}$-bounds.
Relevant for $C_{9,10}$ are $(\delta^d_{23})_{LL}$ and $(\delta^d_{23})_{LR}$, which enter through gluino loops, and $(\delta^u_{23})_{LL}$ and $(\delta^u_{23})_{LR}$, which appear in chargino loops.
Of these, the ones which are not yet very much constrained by bounds on other Wilson coefficients ($C_7$ in this case) and at the same time give a substantial contribution to $C_{9,10}$ are most interesting:
\begin{itemize}
 \item 
The parameter $(\delta^d_{23})_{LR}$ is already tightly constrained by bounds on $C_{7}$ from 
data on the $\bar{B} \to X_s\gamma$ branching ratio~\cite{Gabbiani:1988rb}.
In addition, $(\delta^d_{23})_{LR}$ contributes to $C_{9,10}$ only in double MI 
diagrams~\cite{Lunghi:1999uk}.
We conclude that it plays no role for our analysis.
\item
The parameter $(\delta^d_{23})_{LL}$ is much less constrained by
$\bar{B} \to X_s\gamma$ and 
$B_s-\bar B_s$ mixing \cite{Ball:2003se} than $(\delta^d_{23})_{LR}$.
On the other hand, $(\delta^d_{23})_{LL}$ has little effect on $C_{9,10}$:
Its effect in the $Z$-penguin gluino loop is suppressed with respect to that in the $\gamma$-penguin by a factor of $m_b^2/m_Z^2$ (as discussed, e.g., in~\cite{Lunghi:1999uk}).
The $\gamma$-penguin in turn does not contribute at all to $C_{10}$, and its contribution to $C_9$ is numerically subleading to those from chargino loops, in particular for squark masses above a $\TeV$. Therefore, the parameter $(\delta^d_{23})_{LL}$ is not of interest to our study.
\item
The coefficients $(\delta^u_{23})_{LL}$ and $(\delta^u_{23})_{LR}$ both give a non-negligible contribution to $C_{7}$, which is, however, by far not as significant as that from $(\delta^d_{23})_{LR}$. The impact of $(\delta^u_{23})_{LX},X=L,R$ contributions
to $B_s -\bar B_s$ mixing is very small and negligible \cite{Ball:2003se}.
As exemplified in Table~\ref{table_example_SUSY} for the SUSY benchmark point given in Table~\ref{table_SUSY_point}, $(\delta^u_{23})_{LL}$ gives a much larger contribution to
$C_{7}$ than $(\delta^u_{23})_{LR}$.
Moreover, $C_{10}$ is about an order of magnitude more sensitive to $(\delta^u_{23})_{LR}$ than to $(\delta^u_{23})_{LL}$.
We conclude that $(\delta^u_{23})_{LR}$ is the most relevant parameter for our analysis
on updated constraints from $C_{9,10}$-bounds. 
\end{itemize}
\begin{table}
\centering
\begin{tabular}{|c|c|c|c|c|c|c|c|c|} \hline
$m_{H^\pm}$ &
$\tan\beta$ &
$M_2$ &
$\mu$ &
$m_{\tilde{t}_R}$ &
$m_{\tilde{q}}$ &
$A_t$ &
$m_{\tilde{\nu}}$ &
$m_{\tilde{g}}$
\\\hline
%%%%%%%%%%%%%%%%
300 &
4 &
150 &
$-300$ &
300 &
1000 &
1000 &
100 &
700
\\\hline
\end{tabular}
\caption{Example SUSY point at $\mu_0= 120$ GeV. All masses are 
in GeV. $m_{\tilde{q}}$, $m_{\tilde{t}_R}$ are the diagonal elements of the squark mass matrix as given in the Appendix.
The point corresponds to the following spectrum: 
$m_{\tilde t_1}=236$ GeV,   $m_{\tilde t_2}=1017$ GeV, $m_{\tilde \chi_1}=150$ GeV,   $m_{\tilde \chi_2}=321$ GeV 
and $m_{h^0}=117$ GeV.
\label{table_SUSY_point} }
\end{table}
\begin{table}
\centering
\begin{tabular}{|c|c|c|} \hline
$C_7^{\mathrm{MI},\tilde{\chi}}(\mu_0)$ &
$C_9^{\mathrm{MI},\tilde{\chi}}(\mu_0)$ &
$C_{10}^{\mathrm{MI},\tilde{\chi}}(\mu_0) $ \\\hline
$ 0.01  \duLR -0.38 \duLL$ &
$ 0.17  \duLR -0.11 \duLL$ &
$-2.24  \duLR +0.19 \duLL$ \\
\hline
$C_7^{\mathrm{MI},\tilde{g}}(\mu_0)$ &
$C_9^{\mathrm{MI},\tilde{g}}(\mu_0)$ &
$C_{10}^{\mathrm{MI},\tilde{g}}(\mu_0) $ \\\hline
$ 16.35 \, \ddLR  \,-0.02 \ddLL$ &
$                0.04 \, \ddLL$ & --
  \\
\hline
\end{tabular}
\caption{Flavor off-diagonal Wilson coefficients
at $\mu_0= 120$ GeV for the SUSY point
given in Table~\ref{table_SUSY_point}. Double mass insertions are not shown.
\label{table_example_SUSY} }
\end{table}

For completeness and because of disagreement with the literature in the photon-penguin \cite{Lunghi:1999uk}, see Appendix, we give here
the dependence of $C_{7,9,10}$ on $\duLR$  in the MI approximation,
\begin{align}
C_{7}^{\textrm{MI}, \tilde \chi}(\mu_0)&=
\frac{V_{cs}^*}{V_{ts}^*}  \frac{\lambda_t}{g_2}  \frac{m_W^2}{m_{\tilde q}^2}  F \times   \duLR \, ,
\nn \\
C_{9}^{\textrm{MI}, \tilde \chi}(\mu_0)&= 
\frac{V_{cs}^*}{V_{ts}^*}  \frac{1}{4 \, \sw}
 \frac{\lambda_t}{g_2}
\Big(( 4 \sw-1) F^{Z\textrm{-p}} + 4 \sw \frac{m_W^2}{m_{\tilde q}^2} F^{\gamma \textrm{-p}} -  \frac{m_W^2}{m_{\tilde q}^2} F^{\textrm{box}}\Big) \duLR  \, ,\nonumber
\\
C_{10}^{\textrm{MI}, \tilde \chi}(\mu_0)&=  
\frac{V_{cs}^*}{V_{ts}^*} \frac{1}{4 \, \sw}
\frac{\lambda_t}{g_2}
\Big( F^{Z\textrm{-p}} +   \frac{m_W^2}{m_{\tilde q}^2} F^{\textrm{box}}\Big) \duLR \, , \label{c9c10} 
\end{align}
where $\lambda_t$ denotes the top Yukawa coupling and $s_W^2$ the sine squared of the weak mixing angle.
The expressions $F$ and $F^{\gamma \textrm{-p}}$ are due to $\gamma$-penguin-diagrams, while $F^{Z\textrm{-p}}$ and $F^{\textrm{box}}$ are due to $Z$-penguin- and box-diagrams, respectively.
Their explicit form can be found in the Appendix.
The up-squarks are assumed to be roughly degenerate, except for the right-handed stop, which is allowed to be significantly lighter.
Notice that the $\gamma$-penguin and box-contributions are parametrically suppressed by $m_W^2/m_{\tilde q}^2$ with respect to the $Z$-penguin effect.
Furthermore, since the $Z$-boson mainly has an axial-vector coupling to charged leptons, the $Z$-penguin contribution to $C_9$ is suppressed to the one to $C_{10}$ by $|4 \sw-1| \ll1$ \cite{Buchalla:2000sk}.

\subsection{Numerical analysis \label{sec:numerical}}

For the numerical analysis 
we scan the SUSY parameter space within the ranges given in~Table~\ref{table_SUSYInput}.
All parameters are assumed to be at the electroweak scale, taken to be $\mu_0=120$ GeV.
We fix the lightest sneutrino mass $m_{\tilde{\nu}}=100$ GeV because the dependence of $C_{9,10}^{\rm NP}$ on $m_{\tilde \nu}$ is very mild only.
We denote by $m_{\tilde q}^2$ the $k$th, $k=1, \ldots, 5$ diagonal element of the
up-squark mass matrix and by $m_{\tilde t_R}^2$ the remaining diagonal entry connected
to the right-handed stop, including the $F$ and $D$ terms, see Appendix.
We further fix $m_{\tilde{q}}=1000$ GeV because the bounds
can get only weaker for larger average squark masses.
All  flavor-violating MI parameters except for $\duLR$ are set to zero. 

Then, for each parameter point we require the following constraints to be satisfied:
\begin{itemize}
\item[\it i)] The bounds from the $\bar{B} \to X_s \gamma$ branching ratio: $0.3  \leq  |C_7(\mu_b)| \leq 0.4 $ \cite{Bobeth:2011nj}, see text.
\item[\it ii)] The lightest chargino mass limit: $m_{\chi_1^{\pm}} \geq 94$ GeV  \cite{Nakamura:2010zzi}.
\item[\it iii)] The lightest stop mass limit  $m_{\tilde{t}_1} \geq 100$ GeV from $\mbox{D}\O$\cite{Abazov:2008rc}.
We choose this  lower bound since the stronger one from  CDF, $m_{\tilde{t}_1} \geq 180 $ GeV \cite{Aaltonen:2010uf},  assumes the stop predominantly decaying via charged currents to $b$-quarks, a lepton  and missing energy. The recent LHC findings \cite{ATLAS-CONF-2012-036,CMS-PAS-SUS-11-020} are model-dependent, too, and not taken into account here.
\item[\it iv)] The Higgs mass limits: $m_{h^0} \geq 114.4$ GeV \cite{Nakamura:2010zzi} and $m_{A^0} \geq 200$ GeV.
We require the latter to ensure that $h^0$ is sufficiently SM-like, see text.
\item[\it v)] Electroweak precision tests: $-0.0007 \leq \Delta \rho \leq 0.0017$ \cite{Nakamura:2010zzi}.
\end{itemize}
\begin{table}[h]
\centering
\begin{tabular}{| l | r r r r r r | r|} \hline
&$\tan\beta$&$m_{H^\pm}$ &$M_2$&$|\mu|$&$m_{\tilde{t}_R}$&$A_t$ & $\duLR$ \\\hline
min.&$3$&$300$&$100$&$80 $&$170$&$-3000$ & $-0.85$\\
max.&$15$&$1000$&$1000$&$1000$&$800$&$3000$ & $0.85$\\
\hline
\end{tabular}
\caption{Ranges of  the SUSY parameters (at $\mu_0 = 120$ GeV) used for the parameter scan, see text.
All masses are in GeV.
We fix $m_{\tilde{\nu}}=100$ GeV and $m_{\tilde{q}}=1000$ GeV. 
\label{table_SUSYInput}}
\end{table}
The MSSM allows for large contributions to $C_7$, which can flip its sign while still being
in agreement with data, see {\it i)}. However, the existence of a zero-crossing
of the forward-backward asymmetry in $\bar B \to \bar K^* \mu^+ \mu^-$, see Eq.~(\ref{eq:zeroLHCB}),
enforces ${\rm{sign}}(C_7 C_9)$ to be SM-like. Since $C_9$ cannot change sign within the MSSM \cite{Ali:2002jg}, we allow in our analysis only  MSSM points with SM-like signed
$C_7$.

We use the \texttt{FeynHiggs} code~\cite{Heinemeyer:1998yj,Heinemeyer:1998np,Degrassi:2002fi,Frank:2006yh} v.2.9.0-beta\footnote{We are grateful to Sven Heinemeyer for   \texttt{FeynHiggs}  support  and provision of version 2.9.0-beta.} to check {\it iv)} and {\it v)}. 
  Further, we add 3 GeV to the prediction for $m_{h^0}$ in order to account for the theoretical uncertainty \cite{Degrassi:2002fi}. 
Note further that the recent 95\% C.L. exclusion limits for a SM-like Higgs boson
from ATLAS and CMS suggest a valid range around $(125 \pm \mbox{few} $)  GeV  \cite{:2012si,Chatrchyan:2012tx}.
We checked that while such  values require typically larger $A_t$ than  the constraint {\it iv)}, the generic features of our flavor analysis hold.

Throughout this work we choose the normalization
$\duLR = \left(\Delta_{23}^u\right)_{LR} / \overline M_u^2$, where $\overline M_u^2=\big(  \frac{5}{6}  m^2_{\tilde{q}} + \frac{1}{6} m^2_{\tilde{t}_R} \big)$  to connect the dimensionless MI parameter $\duLR$  to  the off-diagonal element $\left(\Delta_{23}^u\right)_{LR}$ of the up-squark mass matrix in the super-CKM basis.
Note that values of
$\duLR$ outside the range given in Table~\ref{table_SUSYInput} would yield tachyonic squarks due to the large, order one off-diagonal entry.
Note also that a normalization to the  geometric mean of the diagonal entries would result in  increasingly larger values of $\duLR$ for lighter right-handed stops $m_{\tilde{t}_R} < m_{\tilde{q}}$.

We begin by assuming flavor-diagonal SUSY contributions only,  i.e., for $\duLR=0$. In this case
charged Higgs boson and CKM-driven chargino loops  give non-vanishing NP contributions to the Wilson coefficients.
Defining the ratios
\begin{align} R_i \equiv \left| \frac{C_i^{\mathrm{NP}} }{ C_i^{\mathrm{SM}} }\right| 
\end{align}
we find the following, maximally possible NP effects 
\begin{align}
R_9(\mu_b)&\lesssim 3 \%, & R_{10}(\mu_b)&\lesssim 11 \%\, . ~~~~~~~~~~~ (\mbox{MFV}) 
\end{align}
As expected \cite{Ali:2002jg}, the flavor-diagonal SUSY effects on $C_{9,10}$ are small and well within what is allowed by present and near-future data.

Switching on squark flavor violation, the 
possible size of NP effects in the Wilson coefficients increases 
\begin{align} \label{eq:oldbounds}
R_9(\mu_b) &\lesssim 4  \%, & R_{10}(\mu_b)&\lesssim 47 \% \, ,~~~~~~~~~~ (\mbox{no s.l.~bounds})
\end{align}
in particular in $C_{10}$.
Here, the new semileptonic bounds are not yet taken into account. 
Note that the Higgs mass bound {\it iv)} is efficient, as previously noted in~\cite{AranaCatania:2011ak}; if ignored, the allowed range for 
$R_{10}(\mu_b)$ would be more than a factor of two larger.
\begin{figure}[h]  
\begin{center}
\includegraphics[width=0.6\textwidth]{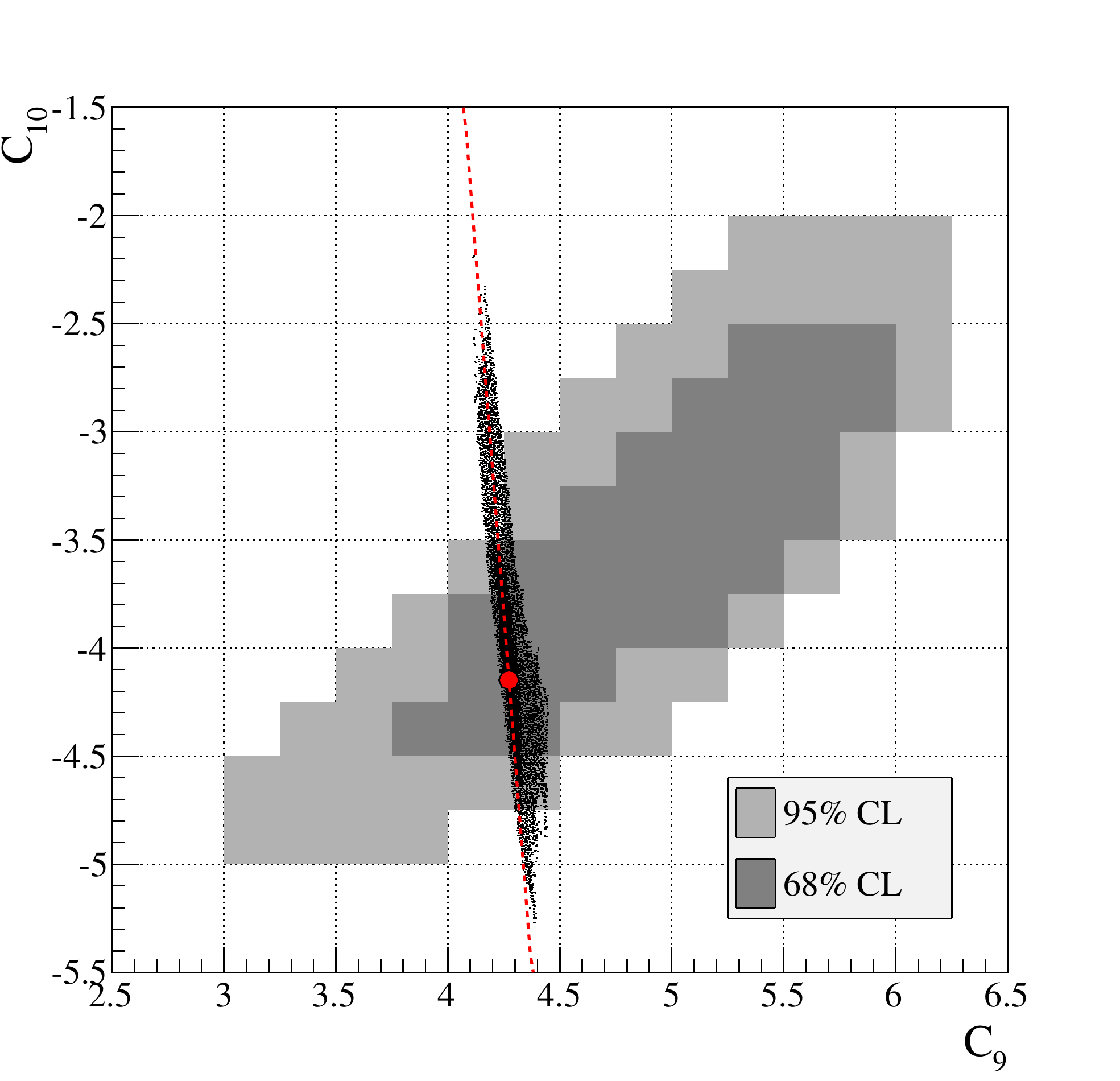} 
\caption{Reach of SUSY models with $\duLR\neq0$ in the $C_9(\mu_b)$--$C_{10}(\mu_b)$ plane for  $C_7(\mu_b) < 0$.
The light (dark) gray shaded areas are the 95\% (68\%) confidence limit bounds which were obtained from $B \to K^{(*)} l^+ l^-$ data in Fig.~7 of Ref.~\cite{Bobeth:2011nj}. The red dotted line denotes the $Z$-penguin correlation 
$C^{Z-\rm p}_{10}/C^{Z-\rm p}_9 =1/(4s_W^2-1)$. The SM point  $(C_9^{\rm SM},C_{10}^{\rm SM})$ is marked by the red dot.
\label{plot-mia-delta23uLR-c7-sm} } 
\end{center}
\end{figure}

The outcome of the full scan is shown in Fig.~\ref{plot-mia-delta23uLR-c7-sm}: 
SUSY points that pass {\it i) - v)} are shown together with the allowed regions (gray areas) in the 
$C_9(\mu_b)-C_{10}(\mu_b)$ plane  from the recent analysis of $\bar B \to \bar K^{(*)} l^+ l^-$ decays \cite{Bobeth:2011nj} for SM-like signed $C_7$.  The SUSY flavor effects are dominated by the $Z$-penguin contribution, correlating $C_9$ and $C_{10}$ as $C^{Z-\rm p}_{10}/C^{Z-\rm p}_9 = 1/(4s_W^2-1)$, see Eq.~(\ref{c9c10}). The latter correlation is shown as the dotted line and is clearly  featured by the model. The analogous solutions with flipped-sign $C_7>0$, which are allowed model-independently  \cite{Bobeth:2011nj}, are excluded in the MSSM (and not shown):
firstly because the $Z$-penguin dominated scatter points miss in this case the allowed $C_9(\mu_b)-C_{10}(\mu_b)$ parameter space and
secondly because of the measurement Eq.~(\ref{eq:zeroLHCB}). 

Including the semileptonic decay data  at  68\%  (95\%) C.L.~we obtain
\begin{align} \label{eq:newbounds}
R_9(\mu_b) &\lesssim 4\% ~(4\%), & R_{10}(\mu_b)&\lesssim 16\% ~(28 \%)  \, ,
\end{align}
cutting into the models' parameter space, cf.~Eq.~(\ref{eq:oldbounds}).
Due to the $Z$-penguin dominance of the SUSY flavor contributions  to $C_9$ and $C_{10}$,
the constraints Eq.~(\ref{eq:newbounds}) are much stronger than those on model-independent scenarios with general $C_9$ and $C_{10}$ \cite{Bobeth:2011nj}.

\subsection{Constraining \texorpdfstring{$(\delta^u_{23})_{LR}$}{delta(u23LR)}\label{sec:bounds}}

\begin{figure}[h] 
\begin{minipage}{0.49\linewidth}
\begin{center}
\includegraphics[width=\textwidth]{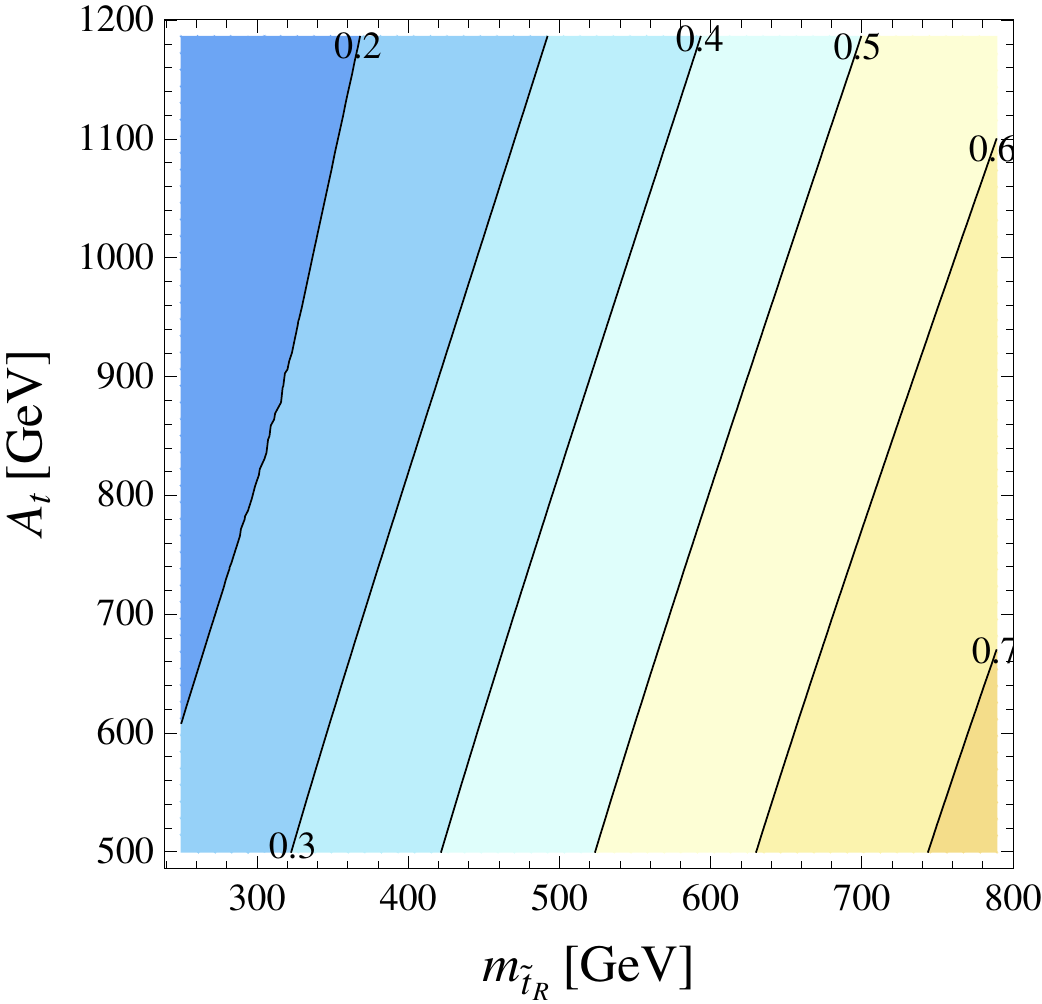}
\end{center}
\end{minipage}
\hfill
\begin{minipage}{0.49\linewidth}
\begin{center}
\includegraphics[width=\textwidth]{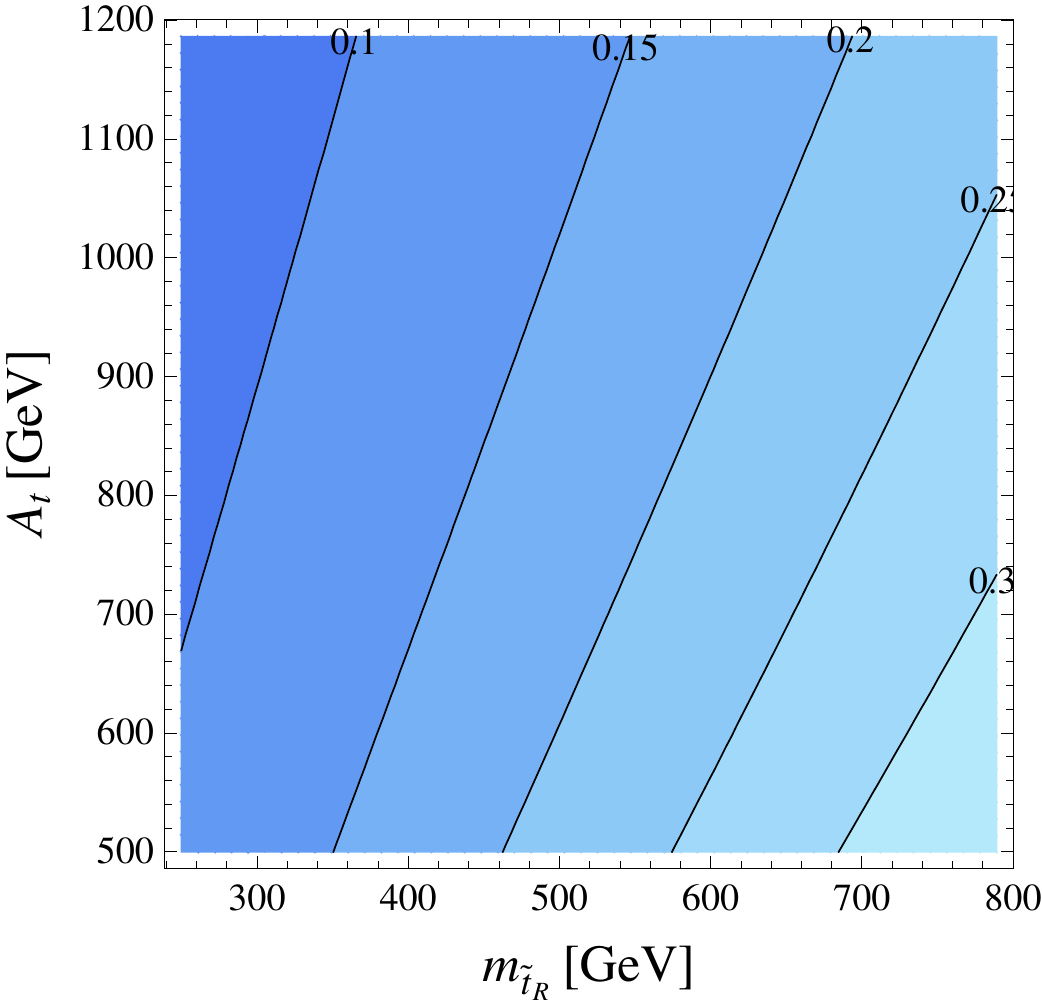}
\end{center}
\end{minipage}
\caption{Upper bounds on $\vert\left(\delta_{23}^u\right)_{LR}\vert$ around the SUSY parameter point defined in Table~\ref{table_SUSY_point}.
Left-hand side: without semileptonic bounds. Right-hand side: including the semileptonic bounds at  68\% C.L.
\label{plot-delta23uLR-boundAgainst-mstR-At-683}
}
\end{figure}
\begin{figure}[h] 
\begin{minipage}{0.49\linewidth}
\begin{center}
\includegraphics[width=\textwidth]{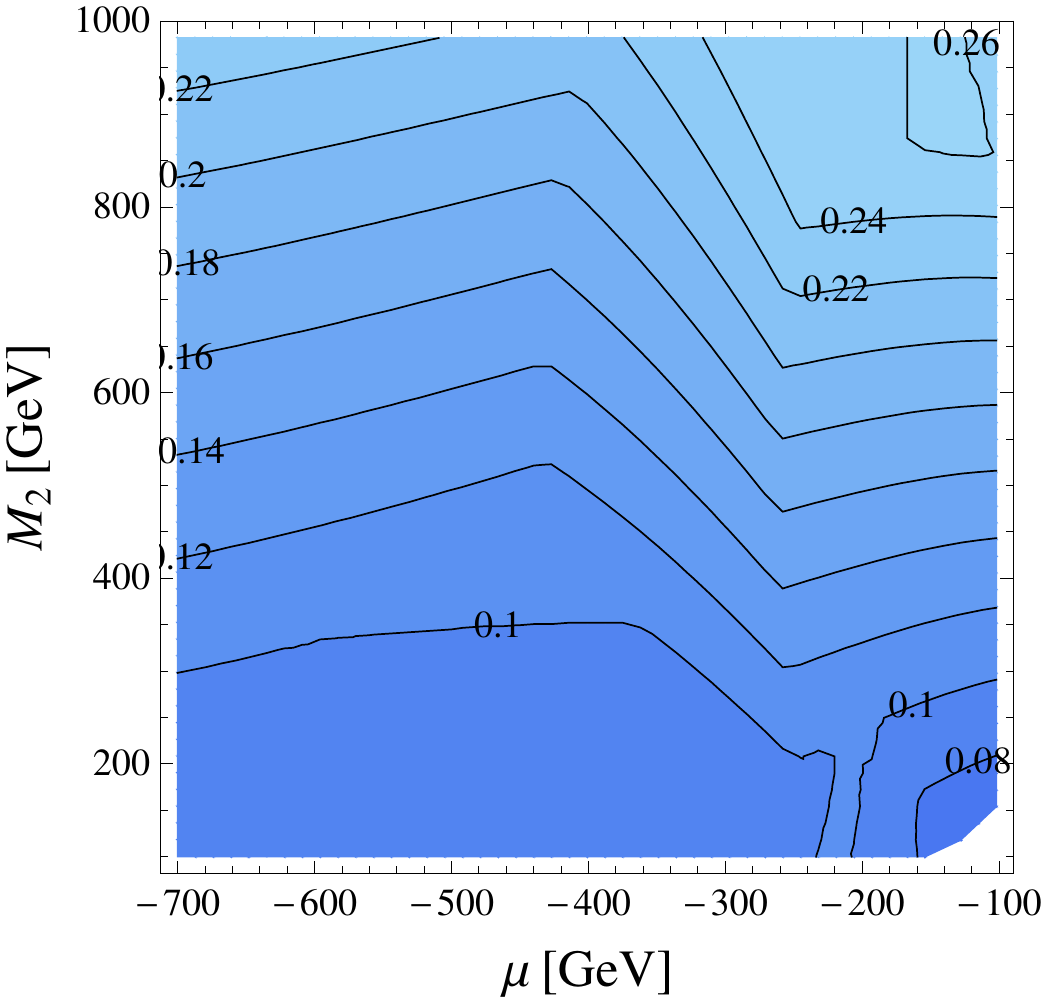}
\end{center}
\end{minipage}
\hfill
\begin{minipage}{0.49\linewidth}
\begin{center}
\includegraphics[width=\textwidth]{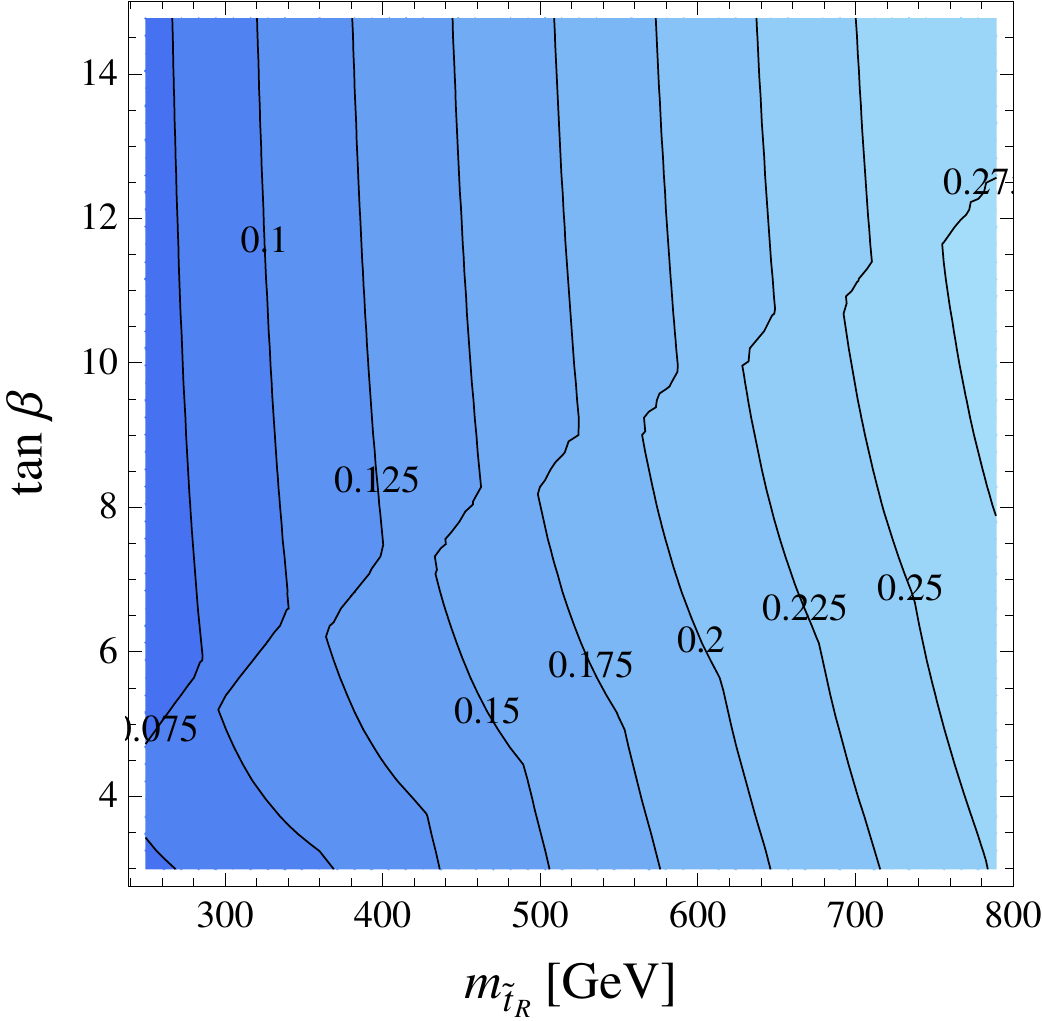}
\end{center}
\end{minipage}
\caption{Upper bounds on $\vert\left(\delta_{23}^u\right)_{LR}\vert$ around the SUSY  point given in Table~\ref{table_SUSY_point}, varying in different directions around the example SUSY parameter point.
Both plots include the 68\% C.L. semileptonic bounds. 
\label{plot-delta23uLR-bound-otherparameters}
}
\end{figure}
Previous works found that  $\left(\delta_{23}^u\right)_{LR}$ is essentially unconstrained by  
$|\Delta B|=|\Delta S|=1$ decay data \cite{Xiao:2006gu}.
We illustrate the impact of the new constraints from semileptonic decays  on squark flavor mixing. In Fig.~\ref{plot-delta23uLR-boundAgainst-mstR-At-683} we show the upper limit on 
$\vert\left(\delta_{23}^u\right)_{LR}\vert$, without (left-hand plot) and including (right-hand plot) the recent data on semileptonic decays. 
For the latter plot we employ the 68\% C.L. bounds that are depicted in dark gray in Fig.~\ref{plot-mia-delta23uLR-c7-sm}.
For definiteness, we choose  the $A_t$-$m_{\tilde{t}_R}$ plane around the SUSY parameter point defined in Table~\ref{table_SUSY_point}.
One observes that, at least in this part of the SUSY parameter space, the bound on $\duLR$ is significantly improved by the new data. 

In Fig.~\ref{plot-delta23uLR-bound-otherparameters} we show the upper limits on $\vert\left(\delta_{23}^u\right)_{LR}\vert$ in the $\mu$-$M_2$ and the $m_{\tilde{t}_R}$-$\tan\beta$ plane, in both cases including the data from $\bar B \to \bar K^{(*)} l^+ l^-$.  
The bounds grow stronger for decreasing $M_2$ and decreasing $m_{\tilde{t}_R}$.
Note that the Higgs mass limits and bounds by electroweak precision tests are discarded in Figs.~\ref{plot-delta23uLR-boundAgainst-mstR-At-683} and \ref{plot-delta23uLR-bound-otherparameters}, as we only want to show here the parametric dependence of the new semileptonic bounds.

Note that \duLR~is also constrained by demanding vacuum stability,
\be
\duLR \leq \frac{m_t}{\overline M_u^2}  \sqrt{2 \overline M_u^2 + \overline M_l^2},
\ee
where $\overline M_u^2$ and 
$\overline M_l^2$ denote an averaged up-squark and slepton mass squared, respectively~\cite{Casas:1996de}.
Requiring only metastability, these bounds are in general weakened but only very little so in case of \duLR~\cite{Park:2010wf}.
Depending on the flavor-diagonal MSSM parameters, the bounds on~\duLR~from $\bar{B} \rightarrow \bar{K}^{(*)} l^+ l^-$ data which we obtain are 
stronger than the vacuum stability bounds. 
For instance, setting $m_{\tilde{q}} = m_{\tilde{l}} = 1$~TeV and $m_{\tilde{t}_R} = 300$ GeV, 
vacuum stability requires $\left(\delta_{23}^u\right)_{LR} \lesssim 0.3$, which is comparable to the bounds shown in the left-hand plot of Fig.~\ref{plot-delta23uLR-boundAgainst-mstR-At-683} but weaker than the ones in the right-hand plot, which includes the new semileptonic rare decay data.

\section{Predictions \label{sec:predictions}}

We present implications for $b$-physics in Section \ref{sec:b}, for RFV models in
Section \ref{sec:RFV} and for rare decays of the top in Section \ref{raretop}.

\subsection{\texorpdfstring{$b$}{b}-physics  \label{sec:b}}
The constraints Eq.~(\ref{eq:newbounds})  imply upper bounds on CP phases of the same order
in the respective Wilson coefficients assuming order one phases in the flavor parameter $\duLR$.
Only little is known to date about the CP-violating phases of $C_{9,10}$: the
relative phase $|\arg C_9 C_{10}^*|$ is constrained by data on $\bar B \to \bar K^* l^+ l^-$ decays at large dilepton masses to be near $\pi$ \cite{Bobeth:2011gi}. The situation will improve 
in the future with more precise data and measurements of CP asymmetries.
Specifically the (naive) T-odd CP asymmetry $\langle A_7^D \rangle$ is unsuppressed by strong phases  \cite{Bobeth:2008ij} and is directly sensitive to CP violation in
$C_{10}$ and hence SUSY flavor. At large dilepton masses the CP  asymmetry
$\langle a_{\rm CP}^{(3)}\rangle$ \cite{Bobeth:2011gi} is promising. The latter is related
to the CP asymmetry of the forward-backward asymmetry put forward in Ref.~\cite{Buchalla:2000sk}.

We predict from Fig.~\ref{plot-mia-delta23uLR-c7-sm}
 that the $\bar B_s \to \mu^+ \mu^-$ branching ratio 
${\cal{B}}(\bar B_s \to \mu^+ \mu^-) \propto f_{B_s}^2 |C_{10}|^2$ is
enhanced (suppressed) with respect to the SM one by at most a factor of 1.3 (0.5) at  95\% C.L. and within the range
    \begin{align} \label{eq:range}
  1  \times 10^{-9} \lesssim  {\cal{B}}(\bar B_s \to \mu^+ \mu^-) < 5 (6) \times 10^{-9} .
     \end{align}
     On the other hand, taking the recent upper limit Eq.~(\ref{eq:bsmumulhcb}) one obtains at 
     95\% C.L.
      \footnote{It has been pointed out recently \cite{deBruyn:2012wk} that the finite lifetime difference in the $B_s$-system causes  ${\cal{B}}(\bar B_s \to \mu^+ \mu^-)$ when extracted from an untagged measurement as in Eq.~(\ref{eq:bsmumulhcb})  to differ from its corresponding value in the unmixed case.  Including the effects from mixing, the limits in Eq.~(\ref{eq:newboundsLHCb}) would get stronger by a few percent.
 }
         \begin{align} \label{eq:newboundsLHCb}
    R_{10}(\mu_b)&\lesssim 46\% ~(20 \%)  \, ,
\end{align}
consistent with Eq.~(\ref{eq:newbounds}). In Eqs.~(\ref{eq:range}) and (\ref{eq:newboundsLHCb}) 
we used for the $B_s$-meson's decay constant 
$f_{Bs}=231(15)(4)\,  \mbox{MeV}$ from Ref.~\cite{Gamiz:2009ku}, corresponding to a SM branching ratio $ {\cal{B}}(\bar{B}_s \to \mu^+\mu^-)_{\rm SM} =(3.1 \pm 0.6) \times 10^{-9}$. The numbers in parentheses correspond to
$f_{Bs}=256(6)(6)\, \mbox{MeV}$ from Ref.~\cite{Simone:2010zz}, with
${\cal{B}}(\bar{B}_s \to \mu^+\mu^-)_{\rm SM} =(3.8 \pm 0.4) \times 10^{-9}$. 
In the lattice results the first
error is statistical and the second one from systematics.

    One observes that the purely leptonic and the semileptonic decays
    give about comparable constraints. Since the former has a larger sensitivity to
    scalar/pseudoscalar operators, which would in SUSY kick in for large values of $\tan \beta$ and a not too heavy Higgs sector and which we neglect, the combined analysis of both modes is most important as they probe complementary NP.

Furthermore, the forward-backward asymmetry in $\bar B \to \bar K^* l^+ l^-$ decays exhibits a 
well-known zero in the SM, roughly determined by $-2 m_b m_B  {\rm Re} (C_7(\mu_b)/C_9(\mu_b))$.
In the SUSY model both $C_7$ and $C_{9}$ are near their respective SM values,
and so is the location of the zero. This is consistent with data, cf.~Eq.~(\ref{eq:zeroLHCB}).

%%%%%%%%%%%%%%%%%%%%%%%%%%%%%%%%%%%%%%%%%%%%
\subsection{Implications for RFV models \label{sec:RFV}} 

\begin{figure}[h]  
\begin{center}
\begin{minipage}{0.49\textwidth}
\begin{center}
\includegraphics[width=\textwidth]{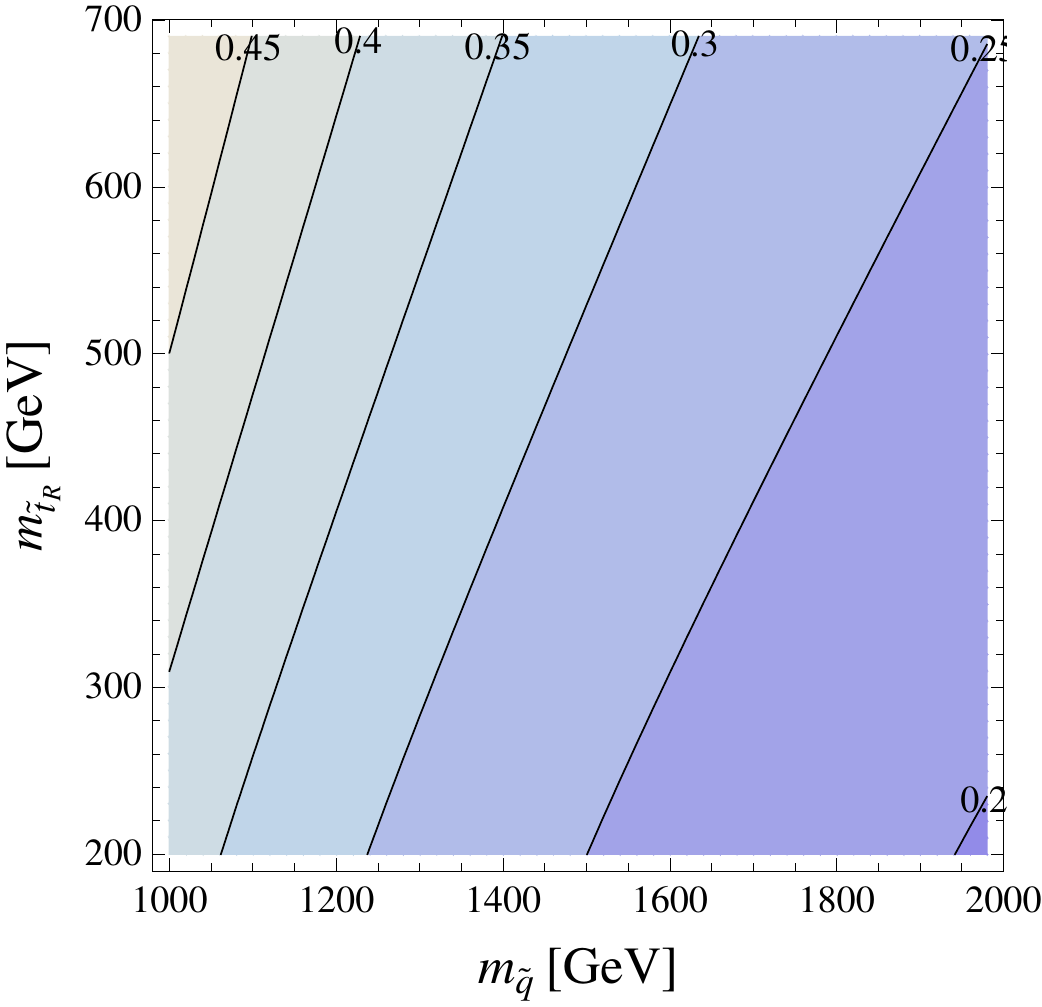}
\end{center}
\end{minipage}
\end{center}
\caption{The required value of  $\duLR$ in the RFV model, see Eq.~(\ref{eq:vcb-rfv}), for
$m_{\tilde g}=1000$ GeV  and $\ddLR=0$.}
\label{plot:deltau23LR}
\end{figure}

As shown in Section \ref{sec:bounds}, 
the current bounds on \duLR~can reach a level of $\sim 0.1$.
This implies constraints
on SUSY flavor models which have rather largish values of  $\duLR$ in this ballpark.
One such model~\cite{Crivellin:2008mq,Crivellin:2011sj} (see references therein for earlier works) is based on radiative flavor violation, where  the small fermion masses and CKM off-diagonal elements originate from quantum loops \cite{Weinberg:1972ws}. 
The CKM matrix is assumed to be the unit matrix at tree level, and the off-diagonal elements are induced by quantum corrections involving off-diagonal trilinear SUSY breaking couplings. 
Specifically, the quark mixing between the second and the third generation is then given as~\cite{Crivellin:2008mq, Crivellin:2011sj}
\begin{align} \label{eq:vcb-rfv}
V_{cb} &= \frac{2 \alpha_s }{3 \pi m_{\tilde{g}}} \bigg( \frac{\left(\Delta_{23}^d\right)_{LR} }{m_b} \widetilde C_0(x,x) -
	\frac{\left(\Delta_{23}^u\right)_{LR} }{m_t} \widetilde C_0(x,y)
    \bigg),
\end{align}
where  $x = m^2_{\tilde{q}}/m^2_{\tilde{g}} $ and $y = m^2_{\tilde{t_R}}/m^2_{\tilde{g}}$ and
\begin{align}
\widetilde{C}_0(x,y) = \frac{(1-y)x\log (x)+(x-1) y \log (y)}{(x-1) (y-1) (x-y)} .
\end{align}
The dimensionless loop function
$\widetilde C_0$ satisfies $\widetilde C_0(1,1)=-1/2$ and $\widetilde C_0(1,0)=-1$.
Above, we allowed for an admixture of flavor generation from squark mixing in the down-sector  
$\left(\delta_{23}^d\right)_{LR}  =\left(\Delta_{23}^d\right)_{LR}/m_{\tilde{q}}^2$ as well.

\begin{figure}[h]  
\begin{minipage}{0.49\textwidth}
\begin{center}
\includegraphics[width=\textwidth]{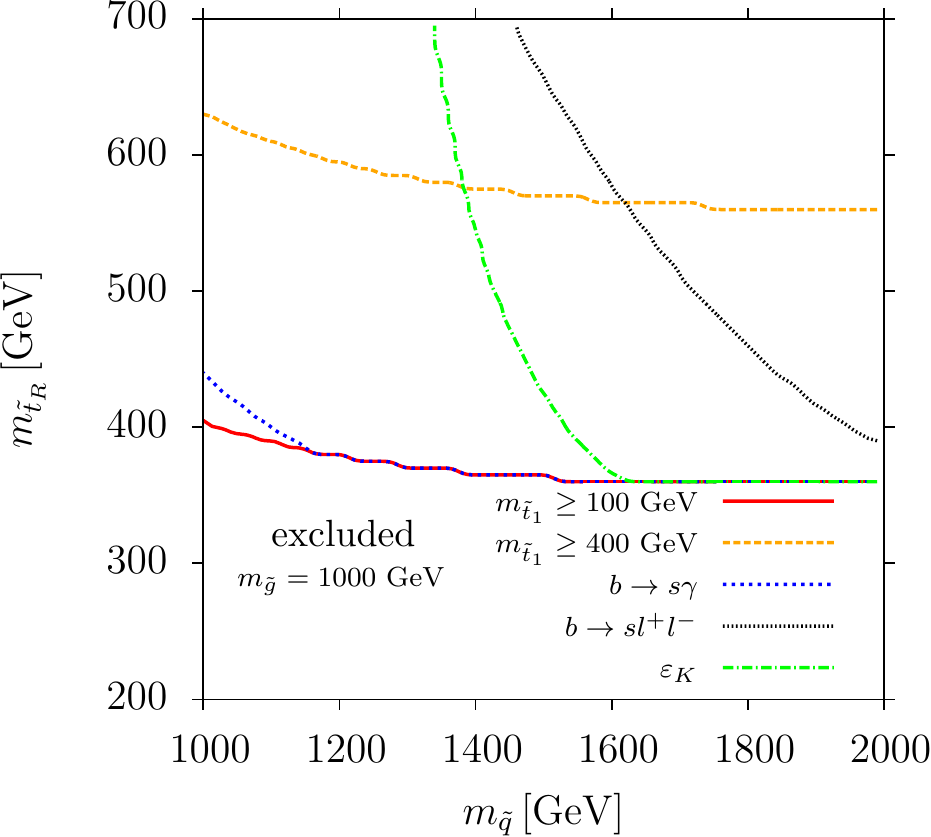}
\end{center}
\end{minipage}
\hfill
\begin{minipage}{0.49\textwidth}
\begin{center}
\includegraphics[width=\textwidth]{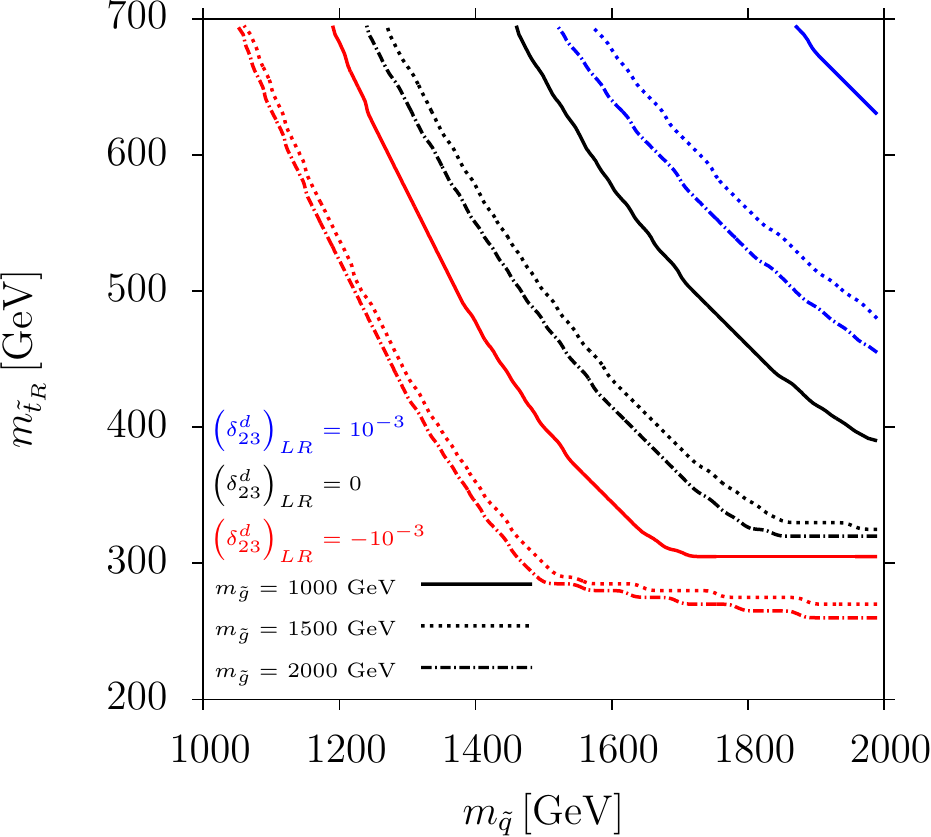}
\end{center}
\end{minipage}
\caption{Excluded regions of RFV parameter space in the $m_{\tilde{q}}-m_{\tilde{t}_R}$ plane 
around the SUSY point given in Table~\ref{table_SUSY_point} with $M_2=800$ GeV.
The left-hand plot displays all bounds including the semileptonic bound at 95\% C.L. (black line), the limit from $m_{\tilde{t_1}} > 100$ GeV (red line), a hypothetical limit $m_{\tilde{t_1}} > 400$ GeV (orange line), $\epsilon_K$ (green line) and $b \to s \gamma$ (blue line) for $m_{\tilde g}=1000$ GeV and $\left(\delta_{23}^d\right)_{LR} = 0$. In the right-hand plot the 95\% C.L. semileptonic bounds are shown for
 different values of $m_{\tilde g}$ and in addition with finite $\left(\delta_{23}^d\right)_{LR} = \pm1\times 10^{-3}$. 
 \label{plot-rfv}}
\end{figure}

{}From Eq.~(\ref{eq:vcb-rfv}) one can determine the flavor mixings
required for a realistic CKM element 
$ V_{cb} = (40.6 \pm 1.3) \times 10^{-3}$ \cite{Nakamura:2010zzi}.
Since $\widetilde C_0<0$, it follows that $\duLR >0$ is required, and that  an additional contribution from
$\ddLR >0\, (<0)$ demands  a larger (smaller) value for  $\duLR$. 
The value of  $\duLR$ required to generate $V_{cb}$ is shown in Fig.~\ref{plot:deltau23LR}  for $m_{\tilde g}=1000$ GeV and $\ddLR=0$.
We evaluate  Eq.~(\ref{eq:vcb-rfv})  at $\mu_0 = 120$ GeV. Since the
requisite $\duLR$ is sufficiently below one, at least for the region displayed and for reasonable gluino masses, the use of the mass insertion approximation is
justified, see also \cite{Crivellin:2011sj}. 
For the gluino-down-squark contribution to the Wilson coefficients we use the mass insertion approximation of the exact results  given in~\cite{Cho:1996we}.

We further evaluate the constraints  from $K^0-\bar K^0$-mixing via up-squark mixing with the third generation, induced by $(\delta^u_{23})_{LR}^* (\delta^u_{13})_{LR}$. The latter factor has a large CP-phase since $(\delta^u_{13})_{LR}$ generates $V_{ub}$  through the analogue of Eq.~(\ref{eq:vcb-rfv}) \cite{Crivellin:2011sj}. 
We use the full, non-MI result given in Ref.~\cite{Colangelo:1998pm} with erratum \cite{Buras:1999da}
and apply $\epsilon_K^{\rm RFV} <0.6 \epsilon_K^{\rm exp}$.\footnote{We thank Andreas Crivellin for
numerical checks  of the chargino contribution to $K^0-\bar K^0$-mixing.}
We use the NLO-RG factor $\eta \simeq 0.8$ \cite{Buras:2001ra,Ciuchini:1998ix} for  the leading $\Delta F=2$  operator $[\bar s \gamma_\mu (1-\gamma_5) d]^2$ with bag parameter $B_K^{\overline{\rm MS}}(2 \, \mbox{GeV})=0.52$ \cite{Colangelo:2010et}.

The interplay of the various constraints is illustrated in Fig.~\ref{plot-rfv}.
We show the exclusion regions  in the $m_{\tilde{t}_R}$-$m_{\tilde{q}}$ plane with SUSY parameters chosen around the SUSY point defined in Table~\ref{table_SUSY_point} with
$M_2=800$ GeV.
Note that in the chargino-loop $b \to s l^+ l^-$ amplitude some CKM elements of the Wilson coefficients, see, e.g., Eq.~(\ref{c9c10}) in the MI approximation, cancel 
against those in Eq.~(\ref{eq:Heff}) to  only diagonal elements $V_{tb} V_{cs}^*$.  For those  we use the physical CKM matrix elements instead of the bare ones \cite{Crivellin:2011sj}.
We find that the constraints from $\epsilon_K$ are weaker than
the semileptonic bounds for not too small values of $M_2$ and for stops sufficiently split from the other squarks, around the point Table~\ref{table_SUSY_point}  for $M_2 \gtrsim 500$ GeV and $m_{\tilde t_R} \lesssim 800$ GeV. In other words, the Glashow-Iliopoulos-Maiani (GIM) suppression can efficiently
make evade the semileptonic bounds.
We learn that the new $\bar{B} \rightarrow \bar{K}^{(*)} l^+ l^-$ data allows to exclude an additional part of the parameter space. The excluded region is larger for lighter gluinos and for contributions from
$\ddLR >0$.

%%%%%%%%%%%%%%%%%%%%%%%%%%%%%%%%%%%%%%%%%%%%
\subsection{Implications for rare top decays \label{raretop}}

Since the same flavor-changing squark mass parameters enter $B$-meson- and top quark decay amplitudes, the improved constraints on $\duLR$ obtained in Section \ref{sec:squark_flavor} can lead to sharper upper bounds on the branching ratio of rare top decays in the MSSM.
Specifically, we consider $t \to c \, V$, where $V$ is either a photon, gluon or a $Z$-boson.
In the SM, the branching ratios are negligibly small, related to a GIM suppression~\cite{AguilarSaavedra:2002ns,AguilarSaavedra:2004wm}. 
Estimates together with upper limits from a
pre-LHC study for the MSSM~\cite{Cao:2007dk} and the requisite branching ratios for $5\sigma$ observations at center-of-mass energy $\sqrt{s}=14$ TeV
with $10\, \mathrm{fb}^{-1}$ and $100\, \mathrm{fb}^{-1}$
of collision data
\cite{Veloso:2008zza} are compiled in Table  \ref{tab:top}.
Thus, an observation of these decays at the LHC is already excluded for the case of the MSSM.
In order to further strengthen this point, we nevertheless find it useful to provide updated bounds on the achievable branching ratios in the MSSM.

\begin{table}
\begin{center}
\begin{tabular}{|c|c|c|c|c|} \hline
                                & SM  &  MSSM pre-LHC &  ATLAS 
$10\, \mathrm{fb}^{-1}$ & ATLAS $100\, \mathrm{fb}^{-1}$
\\\hline
$t \rightarrow c \gamma$ & $4.6 \times 10^{-14}$ & $5.2\times 10^{-7}$   & 
$9.4\times10^{-5}$ & $3.0\times 10^{-5}$\\\hline
$t \rightarrow c g$      & $4.6 \times10^{-12}$ & $3.2\times10^{-5}$   & 
$4.3\times 10^{-3}$ & $1.4\times10^{-3}$\\\hline
$t \rightarrow c Z$      & $1\times 10^{-14}$    & $1.8 \times 10^{-6}$  & 
$4.4\times 10^{-4}$ & $1.4\times 10^{-4}$\\\hline
\end{tabular}
\caption{The branching ratios of rare top decays in the SM \cite{AguilarSaavedra:2002ns,AguilarSaavedra:2004wm} and upper limits in the MSSM from the pre-LHC era \cite{Cao:2007dk}.
The last two columns denote the ATLAS sensitivity  ($5 \sigma $ observation) with $10\, \mathrm{fb}^{-1}$ 
and $100\, \mathrm{fb}^{-1}$, respectively  \cite{Veloso:2008zza}.  \label{tab:top}}
\end{center}
\end{table}

The leading contributions arise from squark-gluino loops with non-vanishing $\duLR$ and/or $\duLL$.
It turns out that the largest effect in the rate into all three final states is due to diagrams involving $\duLR$, cf.~\cite{Cao:2007dk}. 
We calculate the branching ratios for each parameter point in the region defined by Table~\ref{table_SUSYInput} which passes the constraints \textit{i)} to \textit{v)}
as well as the constraints on $C_9$ and $C_{10}$ from $\bar B \to \bar K^{(*)} l^+ l^-$. 
The gluino mass is taken to be $m_{\tilde{g}} = 700\,\GeV$.
For heavier gluinos the upper bounds on the branching ratios decrease further.
We employ the formulae for the $t \to c \, V$ decay widths which are found in~\cite{deDivitiis:1997sh}.
The branching ratios are obtained by normalizing $\Gamma(t \to c V)$ to the SM value for the dominant decay mode of the top quark, $\Gamma(t \to bW) = 1.29\,\GeV$  \cite{Nakamura:2010zzi}. 
Note that, as in Section~\ref{sec:squark_flavor}, we do not rely on the MI approximation.

As a result, we find that the maximal branching ratios compatible with the $68\%$ C.L. constraints from the rare $B$-decays  are
\begin{align} \label{eq:topBR}
\mathcal{B}(t \to c\gamma) &\lesssim 2.1 \times 10^{-8} , &
\mathcal{B}(t \to cg)      &\lesssim  7.2 \times 10^{-7} , &
\mathcal{B}(t \to cZ)      &\lesssim 1.0 \times 10^{-7} .
\end{align}
The constraints improve on the ones from a previous pre-LHC study shown in Table \ref{tab:top}
by a factor of 24, 44 and 18 for decays to $\gamma, g$ and $Z$ plus charm, respectively. The improvement is, depending on the parameter point, due to the
$\Delta B=1$ data but also from the improved squark and gluino mass bounds from the LHC. 
Without the  Higgs mass constraint {\it iv)} enforced using   \texttt{FeynHiggs} v.2.9.0-beta, the limits in Eq.~(\ref{eq:topBR})
would be lifted by about ${\cal{O}}(5-8)$.

%%%%%%%%%%%%%%%%%%%%%%%%%%%%%%%%%%%%%%%%%%%%
\section{Summary \label{sec:conclusions}}

Recent progress  in both the theoretical description and the data 
regarding  $\bar{B} \rightarrow \bar{K}^{(*)} l^+ l^-$ decay distributions 
allows to put new constraints on squark flavor violation. While the current sensitivity in $C_{10}$ is about a factor of two away from  MFV physics, beyond-CKM flavor mixing
is strongly constrained, see Fig.~\ref{plot-mia-delta23uLR-c7-sm}.
This concerns prominently left-right flavor mixing between the second and the third generation in the up-sector encoded in the parameter $\duLR$. With its sensitivity in other  rare processes including $B_s-\bar B_s$ mixing and $b \to s\gamma$ modes being  very weak, $\duLR$ has previously been bounded very loosely only.

We obtain -- depending on the flavor-diagonal SUSY parameters -- constraints as low as $\duLR \lesssim 0.1$. This excludes solutions to the flavor problem with  RFV flavor models 
based on flavor generation in the up-sector and sub-TeV spectra.
Models with horizontal flavor symmetries generically predict $\duLR 
\sim V_{cb} (m_t/m_{\tilde q})$, cf. e.g., \cite{Nir:2002ah}
and are about an order of magnitude below the current limits.
Note that in the future $(\delta^u_{23})_{LR}$ 
could be probed in the presence of an appreciable first-third generation mixing
in  $K \to \pi \nu \bar \nu$ decays \cite{Isidori:2006qy} at the NA62 experiment \cite{Spadaro:2011ue}.

The  new flavor constraints  lead to predictions for $b$- and top physics:
\begin{itemize}
\item 
The $\bar B_s \to \mu^+ \mu^-$ branching ratio 
is, assuming no scalar/pseudoscalar contributions, bounded from below, at about $\sim1  \times 10^{-9} $, see Eq.~(\ref{eq:range}). The constraint on the short-distance coupling $C_{10}$ from the experimental upper limit  Eq.~(\ref{eq:bsmumulhcb})  is consistent with the one
obtained from semileptonic decay data.
Both searches should be pursued further as they probe complementary NP.

\item The forward-backward asymmetry in $\bar B \to \bar K^* l^+ l^-$ decays has a zero at low dilepton mass. The position is  near its SM value, consistent with the preliminary 
determination by the LHCb  collaboration, see Eq.~(\ref{eq:zeroLHCB}).
The forward-backward asymmetry of
the recently observed $\bar B_s \to \phi \mu^+ \mu^-$ and $\Lambda_b \to \Lambda  \mu^+ \mu^-$ 
decays shares the same features \cite{Hiller:2001zj}.
\item Top FCNCs $t \to c \gamma, t \to c g, t \to c Z$ have branching ratios below 
$\lesssim  \mbox{few} \times 10^{-8},  10^{-6}$
and $10^{-7}$, respectively, and are too small to be observed at the LHC with foreseeable luminosities.
\end{itemize}

The $|\Delta B|= |\Delta S|=1$ constraints will improve in the near future with  the large $b \to s l^+ l^-$-induced event samples  expected from the LHC experiments. Progress arises from 
better statistics in combination with the availability of additional observables.
We highlight here those sensitive to CP violation in $C_{10}$ -- induced by a complex-valued 
$\duLR$:
\begin{itemize}
\item The (naive) T-odd CP asymmetry $\langle A_7^D \rangle$ \cite{Bobeth:2008ij}  can be up to order ten percent at small dilepton masses \cite{Altmannshofer:2011gn}.

\item The CP asymmetry  $\langle a_{\rm CP}^{(3)}\rangle$ does not require flavor tagging.  At large dilepton masses it can reach a few percent \cite{Bobeth:2011gi}.
\end{itemize}

The flavor constraints reported here get stronger for lighter stops, an ingredient of TeV-scale model building, recently, e.g., \cite{Papucci:2011wy,Csaki:2012fh,Craig:2012yd,Craig:2012di}, with possibilities to be seen directly at the LHC. If realized in nature, chances are that such models or others based on the generic  MSSM with light stops will show up one way or the other, or both.

\bigskip

{\bf Note added:} During the publishing process a related and complementary work\cite{Mahmoudi:2012un}
on MFV SUSY constraints at large $\tan \beta$ from rare $B$-decay data appeared.
%%%%%%%%%%%%%%%%%%%%%%%%%%%%%%%%%%%%%%%%%%%%
\section*{Acknowledgements}

We are happy to thank Christoph Bobeth, Andreas Crivellin and Danny van Dyk for useful exchanges and Thomas Hahn and Sven Heinemeyer for \texttt{FeynHiggs} support.
We are grateful to Gino Isidori for reminding us of the erratum to \cite{Colangelo:1998pm}.
This work is supported in part by the {\it  Bundesministerium f\"ur Bildung und Forschung (BMBF)} and the {\it German-Israeli Foundation for Scientific Research and Development (GIF)}.
C.~G.~acknowledges support by the {\it Swiss National Science Foundation}.
G.H.~gratefully acknowledges the hospitality and
stimulating atmosphere provided by the Aspen Center for Physics where parts of this work have been done.
C.~G.~would like to thank the Physics Department at Boston University, where part
of this work was done, for kind hospitality.

%%%%%%%%%%%%%%%%%%%%%%%%%%%%%%%%%%%%%%%%%%%%
\begin{appendix}

\renewcommand{\theequation}{A.\arabic{equation}}
\setcounter{equation}{0}  

\section{Loop functions}

We derived the dependence of $C_{7,9,10}$ on $\duLR$  in the MI approximation, cf. Eq.~(\ref{c9c10}), from the exact expressions in \cite{Cho:1996we}.
We find the following loop functions:
\bea
F(x_1,x_2,x_{\tilde t_R})& =&  \frac 16 \hat{x}_{av} \sum_{i=1,2} V_{i1} V_{i2}^* \, x_i^2  \frac{f_1(x_i/x_{\tilde
t_R})-f_1(x_i)}{x_i/x_{\tilde t_R}-x_i} ,
\\
F^{Z\textrm{-p}}(\hat x_1,\hat x_2,\hat x_{\tilde t_R})&=& 
\hat{x}_{av} \sum_{i,j=1,2}   V_{j1} V_{i2}^* \Big\{ 
U_{j1}^* U_{i1} 
\sqrt{\hat x_i \hat x_j} 
\frac{
c_0(\hat x_{\tilde t_R}, \hat x_i, \hat x_j) -
c_0 (1, \hat x_i, \hat x_j )
}{
\hat x_{\tilde t_R} - 1
}
\nn \\
&& \hspace{60pt}
-2 V_{j1}^* V_{i1} 
\frac{
c_2(\hat x_{\tilde t_R}, \hat x_i, \hat x_j) -
c_2 (1, \hat x_i, \hat x_j )
}{
\hat x_{\tilde t_R} - 1
}
\nn \\
&&\hspace{60pt}
+ 2 \delta_{ij}
\frac{
c_2(\hat x_j, 1, \hat x_{\tilde t_R}) -
c_2 (\hat x_j, 1, 1 )
}{
\hat x_{\tilde t_R} - 1
}
\Big\} ,
\\
F^{\gamma \textrm{-p}}(\hat x_1, \hat x_2, \hat x_{\tilde t_R})&=&
\frac 19 \hat{x}_{av} \sum_{i=1,2} V_{i1} V_{i2}^* \ \hat x_{\tilde t_R}
 \frac{\hat x_i/\hat x_{\tilde t_R} f_7(\hat x_i/\hat x_{\tilde t_R})-\hat x_i f_7(\hat x_i)}{\hat x_i/\hat x_{\tilde t_R}-\hat x_i} ,
\\
F^{\textrm{box}}(\hat x_1,\hat x_2,\hat x_{\tilde t_R},\hat x_{\tilde \nu_1})&=& 
\!\! 4 \hat{x}_{av}\!\!\!\!  \sum_{i,j=1,2} \! \! V_{i1} V_{i2}^*   |V_{j1}|^2
\frac{
d_2( \hat x_i, \hat x_j,\hat x_{\tilde t_R},\hat x_{\tilde \nu_1})  -
d_2 (\hat x_i, \hat x_j ,1,\hat x_{\tilde \nu_1})
}{
\hat x_{\tilde t_R} - 1
} ,
\eea
where
\be
x_i = 1/\hat x_i = \frac{m^2_{\tilde q}}{m^2_{\chi_i}} ,~( i=1,2) \,,  \quad \quad
x_{\tilde t_R} =1/\hat x_{\tilde t_R} = \frac{m^2_{\tilde q}}{m^2_{\tilde t_R}}\, , \quad
x_{\tilde \nu_1}=1/\hat x_{\tilde \nu_1} = \frac{m^2_{\tilde q}}{m^2_{\tilde \nu_1}}  
\ee
and $\hat{x}_{av} = \frac{1}{6}(5 + \hat{x}_{\tilde{t}_R})$.
The functions $f_i,c_i,d_2$ are  defined in~\cite{Cho:1996we}. 
The unitary matrices $U$ and $V$ which appear above diagonalize the chargino mass matrix:
\be
U^* M_{\tilde{\chi}^\pm}  V^\dagger = \mathrm{diag}\left(m_{\tilde{\chi}_1}, m_{\tilde{\chi}_2} \right)  \,,
\ee
where 
\begin{align}
M_{\tilde{\chi}^\pm} &= \vect{
M_2 & \sqrt{2} m_W \sin\beta \\
\sqrt{2} m_W \cos\beta & \mu 
} \,.
\end{align}
The up-squark mass matrix is parametrized as 
\begin{align}
M^2_{\tilde{u}} &= \vect{
 \left( M^2_{\tilde{u}} \right)_{LL} & \left( M^2_{\tilde{u}} \right)_{LR} \\
 \left( M^2_{\tilde{u}} \right)_{LR}^\dagger & \left( M^2_{\tilde{u}} \right)_{RR}
} 
\end{align}
with $\left( M^2_{\tilde{u}} \right)_{LL} = m^2_{\tilde{q}} \, \mathbbm{1}_{3\times 3}$, 
\begin{align}
\left( M^2_{\tilde{u}} \right)_{LR} &= \vect{
0 & 0 & 0  \\
0 & 0 & \left(\Delta_{23}^u\right)_{LR} \\
0 & 0 & \left(\Delta_{33}^u\right)_{LR}  
}, \qquad
\left( M^2_{\tilde{u}} \right)_{RR} = \vect{
m^2_{\tilde{q}}  & 0 & 0  \\
0 & m^2_{\tilde{q}} & 0  \\
0 & 0 & m^2_{\tilde{t}_R}
},
\end{align}
with the diagonal elements $m^2_{\tilde{q}}$,  
$\left(\Delta_{33}^u\right)_{LR} = m_t \left( A_t - \mu \cot\beta\right)$ and $m^2_{\tilde{t}_R}$, defined as
\begin{align}
m^2_{\tilde{t}_R} &= m^2_{\tilde{t}_R,\,\mathrm{soft}}  + m_t^2+ \frac{2}{3} s^2_W m_Z^2 \cos 2\beta ,
\end{align}
where $\tilde{m}^2_{\tilde{t}_R,\,\mathrm{soft}}$ denotes the pure soft term contribution.
The stop masses are obtained after diagonalization of $M^2_{\tilde{u}}$:
\begin{align}
m^2_{\tilde{t}_{1,2}} = \frac{1}{2} \left( 
	m_{\tilde{q}}^2+m_{\tilde{t}_R}^2 
	\mp \sqrt{\left(m_{\tilde{q}}^2-m_{\tilde{t}_R}^2\right)^2+4  \left(\Delta_{23}^u\right)_{LR}^2 +4  \left(\Delta_{33}^u\right)_{LR}^2 }
	\right) .
\end{align}
The other squark masses are then given as $m_{\tilde{q}}$. 

Our findings for the Wilson coefficients $C_{7,9,10}$ agree with Ref.~\cite{Lunghi:1999uk} if the following modifications are made to the results of Ref.~\cite{Lunghi:1999uk}:
\begin{enumerate}
\item[1.)] Eqs.~(25) and (26), ($\gamma$-penguin with chargino loop, contributions to $C_9$):
$1/3 P_{042}$ has to be replaced by $-1/3 P_{042}$.
\item[2.)] Eq.~(26), ($\gamma$-penguin with chargino loop, contributions to $C_7$):
The complex conjugation of $U_{i2}$ has to be removed.
\item[3.)] 
Eq.~(31), ($\gamma$-penguin with gluino loop, contributions to $C_7$):
The $(\delta_{23}^d)_{LL}$ contribution has to be multiplied by 2 and $(\delta_{23}^d)_{RL} $ has to be replaced by $- (\delta_{23}^d)_{LR}$.
\item[4.)] Eq.~(34), ($\gamma$-penguin with chargino loop, case of light stop, contributions to $C_9$):
The expression has to be multiplied by a factor 1/6.
\end{enumerate}
We explicitly checked the above corrections by direct calculations using the conventions
of  \cite{Cho:1996we}. Note that
the relative minus sign between \cite{Lunghi:1999uk} and \cite{Cho:1996we} in front of 
$(\delta_{23}^d)_{LR}$ in the $\gamma$-penguin  contribution to $C_7$, see item 3.,  can be absorbed by a rephasing of the gluino mass and the $A$ terms, or equivalently, by
rephasing the right-chiral down squarks in the super-CKM basis.
The corrections 1., 2.~and 4.~are in agreement with Ref.~\cite{Gabrielli:2004yi}.

\end{appendix}

%%%%%%%%%%%%%%%%%%%%%%%%%

\end{document}